\documentclass[pre,groupedaddress,showkeys,showpacs,twocolumn]{revtex4}
\usepackage{amsmath}
\usepackage{graphics}
\usepackage{graphicx}
\usepackage{amsfonts}
\usepackage{amssymb}
\usepackage{dcolumn}
\usepackage{bm}
\usepackage{natbib}
\begin{document}
\title{Transient sliding of thin hydrogel films: the role of poroelasticity}
\author{Lola Ciapa}
\author{Jessica Delavoipi\`ere}
\author{Yvette Tran}
\author{Emilie Verneuil}
\author{Antoine Chateauminois}
\email[]{antoine.chateauminois@espci.fr}
\affiliation{Soft Matter Science and Engineering Laboratory (SIMM), CNRS UMR 7615, ESPCI Paris, PSL University, Sorbonne Universit\'e, F-75005 Paris, France.}
\date{\today}
\begin{abstract}
We report on the transient frictional response of contacts between a rigid spherical glass probe and a micrometer-thick poly(dimethylacrylamide) hydrogel film grafted onto a glass substrate when a lateral relative motion is applied to the contact initially at rest. From dedicated experiments with \textit{in situ} contact visualization, both the friction force and the contact size are observed to vary well beyond the occurrence of a full sliding condition at the contact interface. Depending on the imposed velocity and on the static contact time before the motion is initiated, either an overshoot or an undershoot in the friction force is observed. These observations are rationalized by considering that the transient is predominantly driven by the flow of water within the stressed hydrogel networks. From the development of a poroelastic contact model using a thin film approximation, we provide a theoretical description of the main features of the transient. We especially justify the experimental observation that the relaxation of friction force $F_t(t)$ toward steady state is uniquely dictated by the time-dependence of the contact radius $a(t)$, independently on the sliding velocity and on the applied normal load.
\end{abstract}
\maketitle
\section*{Introduction}
The lubricating properties of polymer hydrogels in aqueous environments are of paramount importance in many biological systems (articular cartilages, mucin layers on the surface of cornea,...) and in biomedical engineering (contact lenses,...) where they can provide very low friction. One emerging issue is the ability of such systems to maintain low friction when sliding is initiated after prolonged static (non sliding) periods. Indeed, some studies dealing with cartilages~\cite{yarimitsu2010,ateshian2009,sakai2012,murakami2015} or synthetic hydrogels \cite{kagata2003, dolan2017, reale2017,shoaib2019,han_effect_2018} showed that increasing contact times can result in a significant increase in the static coefficient of friction characterizing the transition from rest to steady-state sliding with potential detrimental effects on the wear resistance. For hydrogels, these so-called stiction phenomena have been recognized to involve complex interplay between physicochemical interactions across the contact interface and fluid transport within the stressed hydrogel network. The latter are commonly called poroelastic flows: they result from the balance between elastic deformation of the polymer network and solvent flows in the permeable structure of the network in response to a load \cite{delavoipiere2016,hui2005}. In experimental conditions where poroelastic transport during sliding was avoided, a peak in the friction force at onset of sliding was measured in various hydrogel systems: polyelectrolyte hydrogels \cite{kagata2003} or cellulose-based hydrogels \cite{dolan2017}. Its amplitude was found independent on sliding velocity and related to molecular interactions across the interface.\\
In more complex experimental situations where spherical probes slide against various hydrogel substrates \cite{reale2017,shoaib2019,han_effect_2018}, both interfacial effects and fluid transport are at stake {\it a priori}. These studies evidence the development of sharp static friction peaks whose magnitude increases with the static contact time preceding the onset of lateral motion. The interpretation is that poroelastic flow during static time induces an increase in both the contact size and in the adhesion of the gel. The latter is not measured directly and instead, is inferred by the authors from measurements of a pull-off force, and attributed to the increase in polymer concentration within the loaded region as a result of the squeeze-out of water. Assumption is made implicitly by the authors that friction results from pinning/depinning mechanisms between polymer chains and the probe, an hypothesis which forms the basis of the classical Schallamach model~ \cite{schallamach1963} for rubber friction. Accordingly, the density of pinning sites, i.e. the friction force, should increase
with polymer concentration. In these descriptions, stiction remains therefore considered essentially as an interfacial phenomenon where the static friction force at the onset of slippage is defined as the product of a time-dependent contact area by a time-dependent adhesion term. \\
Along the same lines, the progressive relaxation of the friction force after the stiction peaks~\cite{han_effect_2018} was analyzed according to previously developed fracture mechanics descriptions of the stiction of adhesive contacts with rubbers~\cite{barquins1975a,savkoor1992,chateauminois2010}. Assumption is made that, upon application of lateral motion, slip is progressively invading the contact from its periphery according to a mechanism which is reminiscent of the propagation of an interface crack. In this model, the occurrence of a stiction peak is interpreted as the consequence of a transition from stable to unstable slip propagation. However, no evidence of heterogeneous slip exists due to the lack of contact images.\\

\indent In this study, we tackle the stiction problem from a different perspective where we consider the viscous dissipation induced by the poroelastic flows within the bulk polymer network instead of adhesive failure at the interface. Recent numerical simulations have indeed demonstrated that the relaxation dynamics of the friction force on porous hydrogel can be accounted for by poroelastic dissipation when the interface is set to be frictionless.~\cite{qi2020} Here, conditions where fluid transport effects dominate over interfacial contributions to the friction were met experimentally by using thin hydrogel layers and rigid substrates. As a result of contact mechanical confinement, we have previously shown that steady-state friction onto thin poly(dimethylacrylamide) (PDMA) gel layers is predominantly driven by poroelastic dissipation~\cite{delavoipiere2018} with negligible contributions of interfacial interactions. \\
Here, we offer to describe the contribution of poroelastic flows to the kinetics of transient friction when rigid spheres slide against thin hydrogel films. We show that the transients extend well beyond the occurrence of full sliding at the contact interface, and that it can be accounted for by poroelastic flow in the contact zone.\\
In a first part, we investigate how the static contact time affects the time-dependent friction force and contact size during transient sliding for various values of the P\'eclet number defined as the ratio of advective to diffusive contribution to the fluid transport rate. We define the conditions where a stiction peak exists in terms of static contact time values and P\'eclet number. 
We further show the existence of a unique, power law, relationship between the friction force and the contact radius whatever the applied contact load and sliding velocity. These results are discussed within the framework of a poroelastic contact model developed in a thin film approximation, where we consider that the friction force arises from dissipation due to the advective component of pore pressure distribution.\\
%
\section*{Experimental section}
%
\begin{figure}[!ht]
	\centering
	\includegraphics[width=1 \columnwidth]{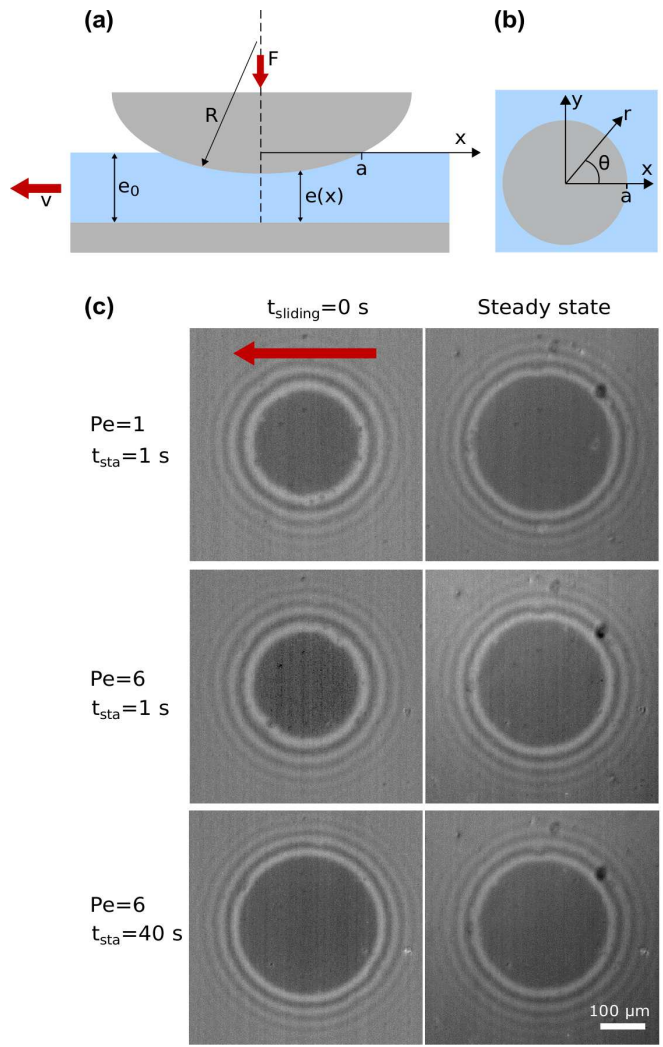}
	\caption{(a) Schematics of a glass sphere (radius $R$) sliding under an imposed load $F$ on an hydrogel layer (thickness $e_0$) immersed in water and grafted to to a glass substrate. A lateral displacement with velocity $v$ is imposed to the substrate after a static contact time $t_{sta}$ as indicated by the arrows. The gel thickness in the contact is $e(x)$. The contact radius measured along the $x$-axis is denoted $a$. (b) Bottom view of the contact. (c) Contact images under sliding ($F=200$~mN) for $Pe=1$ and $Pe=6$, when sliding is initiated (left) and in steady state (right) . Sliding motion is initiated after static contact times $t_{sta}=1$~s or $t_{sta}=40$~s. The hydrogel/glass sphere interface appears as a black zone while rings are interference fringes of equal thickness in white light.}
	\label{fig:schem_images}
\end{figure}

All the experiments to be reported were carried out using poly(dimethylacrylamide) (PDMA) hydrogel films covalently grafted onto glass substrates. These films were synthesized by simultaneously crosslinking and grafting preformed ene-functionalized polymer chains onto glass substrates using a thiol-ene click reaction which is fully described elsewhere~\cite{cholletb2016,cholleta2016,li2015}. This procedure ensured that the films are homogeneously crosslinked through their thickness. In order to achieve good adhesion between the hydrogel film and the glass substrate, we carried out thiol-modification of the borosilicate glass surfaces. The resulting covalent bonding between the film and the glass substrate (through thiol-ene reaction) prevented any interface debonding during swelling and friction. The thickness of the PDMA films was $800 \pm 20$~nm in the dry state, as measured by ellipsometry. Ellipsometry measurements in water provided a swelling ratio of $1.9 \pm 0.1$ for the fully hydrated films, i.e. a thickness $e_0$ of about 1.5~$\mu$m in the fully swollen state.\\
Friction experiments were carried out using a spherical borosilicate glass probe (radius of curvature $R=25.9$~mm) under imposed normal load $F$ (from 40~mN to 600~mN) and driving velocity $v$ (from 1~$\mu$m~s$^{-1}$ to 45~$\mu$m~s$^{-1}$) with the contact fully immersed within deionized water (see Fig.~\ref{fig:schem_images}a). In addition to lateral force measurements with mN accuracy, Reflection Interference Contrast Microscopy (RICM)~\cite{theodoly2009} images of the immersed contact were continuously recorded through the glass substrate using a CMOS camera (600x600 pixels with 12 bits resolution), a combination of crossed-polarizers and quarter-wave plates, and white light illumination. For a detailed description of the used custom-built device, the reader may refer to our previous work.~\cite{delavoipiere2018} From the contact images, the contact shape was measured over time. The contact was found to be circular with a radius $a$ as shown in Fig.~\ref{fig:schem_images}. From the RICM images, the optical contrast measured at the gel/probe interface showed that no lubrication water film is trapped at this interface. This observation is further supported by theoretical considerations on elastohydrodynamic lubrication detailed elsewhere~\cite{delavoipiere2018}.
No damage to the films was evidenced from \textit{in situ} contact visualization. In addition, the contact conditions ensured that the water content $\phi$ of the gel network during sliding was always above the threshold corresponding to the glass transition of PDMA (i.e. $\phi \approx 0.2$ as detailed in Delavoipi\`ere~\textit{et al}~\cite{delavoipiere2016}).\\
In a previous investigation, we have shown from rheology measurements on swollen PDMA films that they are of very low viscoelasticity ($\tan \delta < 0.05$ at 1~Hz). Moreover, the characteristic frequency $v/a$ involved in our friction experiments is less than 10~Hz, which is more than 5 orders of magnitude lower than the estimated glass-transition frequency of the hydrated PDMA network at room temperature. As a consequence, any contribution of viscoelasticity to friction can be neglected.
%
\section*{Experimental results} 
\begin{figure}[!ht]
	\centering
	\includegraphics[width=0.9 \columnwidth]{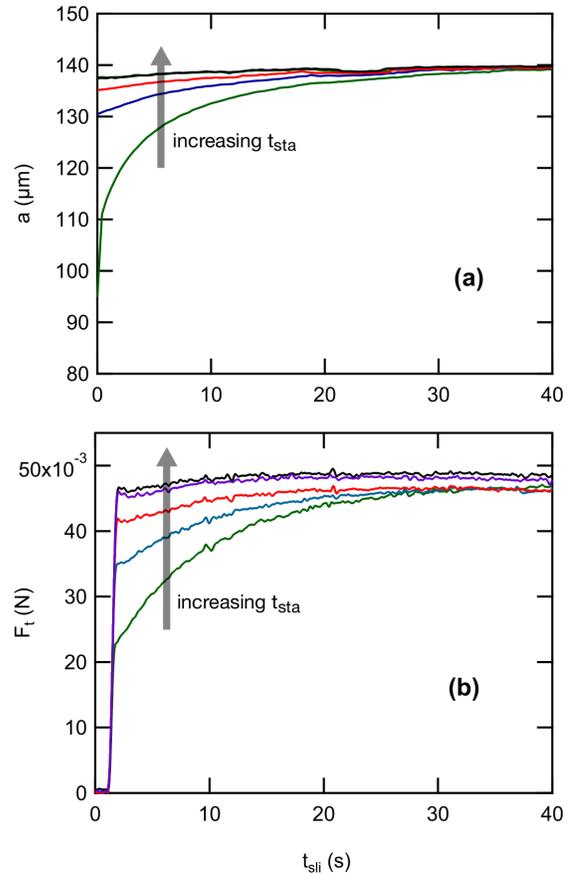}
	\caption{Contact radius $a$ (a) and friction force $F_t$ (b) as a function of sliding time $t_{sli}$ during lateral motion at imposed velocity $v=10 \: \mu$m~s$^{-1}$ and applied normal force $F=200$~mN. Sliding motion is initiated after various static contact time $t_{sta}$. From bottom to top: $t_{sta}=$~1.4, 8.9, 19.1, 38.8 and 58.5~s. For the considered normal load, velocity and poroelastic time ($\tau=19$~s), $Pe=1$.}
	\label{fig:ft_a_ts}
\end{figure}

%
As mentioned above, the objective is to investigate the contribution of poroelastic flow to friction force and contact shape during transient sliding. For that purpose, the P\'eclet number $Pe$ is defined as the ratio of advective to diffusive contributions to the fluid transport rate. Here, advection arises from the sliding motion at velocity $v$ of the contact of radius $a$. The diffusive mechanism is the fluid flow in the hydrogel arising from pore pressure gradients and characterized by the poroelastic time $\tau$. Accordingly, this P\'eclet number can be expressed as
\begin{equation}
Pe=\frac{\tau v}{a}
\end{equation}
As detailed elsewhere~\cite{delavoipiere2016}, the poroelastic time $\tau$ characterizes the indentation kinetics of the sphere into the gel film under static normal load. It involves the elastic and permeation properties of the hydrogel film, its thickness $e_0$ together with the contact conditions (normal load $F$ and radius of curvature $R$ of the spherical probe).\\
\indent In what follows, we first focus on the time-dependence of the friction force $F_t$ in low P\'eclet regime, i.e. $Pe \lesssim 1$. In order to vary the initial swelling state of the hydrogel, we varied the static contact time $t_{sta}$ preceding the onset of the imposed motion. Example images of the contact are shown in Fig. 1c (top row) at the onset of sliding (left) and in steady state (right). In Fig.~\ref{fig:ft_a_ts}, we report data for $F=200$~mN and a sliding velocity $v=10 \: \mu$m~s$^{-1}$ for a series of experiments with static contact times $t_{sta}$ ranging from 2 to 60~s before the imposed lateral motion is initiated. As detailed below, the characteristic poroelastic time for the considered normal load $F$ is $\tau=19$~s. The P\'eclet number is 1.
\begin{figure}[!ht]
	\centering
	\includegraphics[width=0.9 \columnwidth]{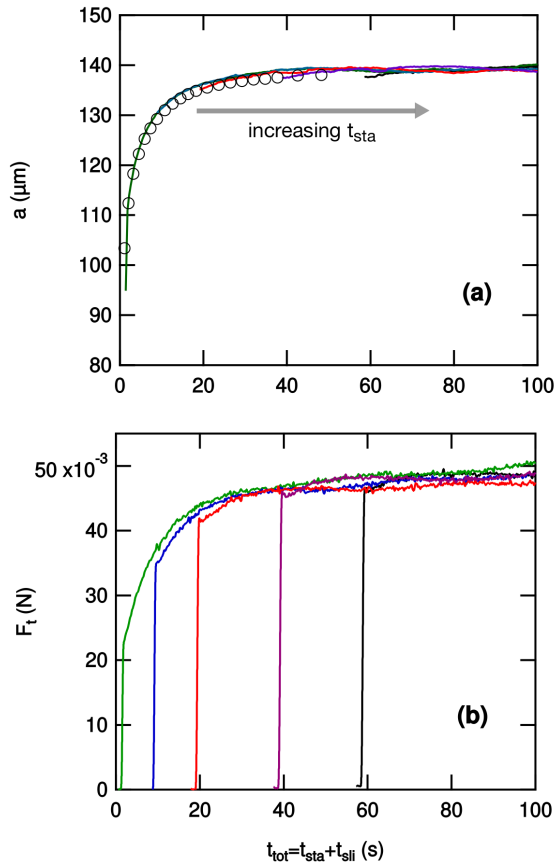}
	\caption{Same results as in Fig.~\ref{fig:ft_a_ts} but as a function of the total contact time $t_{tot}$ defined as the sum of the static contact time $t_{sta}$ and the sliding time $t_{sli}$. In (a), the open symbols correspond to the measured time dependence of the contact radius $a$ under a static normal indentation load only.}
	\label{fig:ft_a_ttot}
\end{figure}
After a sharp initial increase in lateral force from rest, both $F_t$ and $a$ increase toward the same steady-state values within a characteristic time of the order of a few tens of seconds. Here, no peak in $F_t$ is detected. As mentioned in the introduction, classical description of incipient sliding often consider the progressive development of interface slip from the periphery of the contact~\cite{barquins1975a,mindlin1953,chateauminois2010} until a full sliding condition is achieved at the contact interface. Here, the occurrence of such a phenomenon should be restricted to a relative lateral displacement $d$ of order $e_0$, which corresponds to shear deformations $d/e_0$ of order 1 before full slip occurs at the interface. A displacement of a few micrometers corresponding to a fraction of a second for the velocity under consideration, such a full slip condition is therefore achieved during the initial sharp increase in $F_t$. In that respect, the transient regime following the initial increase in $F_t$ in Fig.~\ref{fig:ft_a_ts} does not pertain to the classical descriptions of stiction. Instead, poroelastic flow within the gel layer must be considered.\\
The relevance of such phenomena is evidenced from a consideration of the frictional kinetics as a function of the total contact time defined as the sum of the static contact time and the sliding time. When $F_t(t)$ and $a(t)$ data are reported as a function of this total contact time (Fig.~\ref{fig:ft_a_ttot}), master curves are obtained for all the experiments carried out with varying static contact times. Moreover, the master curve for the contact radius matches the indentation curve $a(t)$ measured in static conditions, i.e. $v=0$ (shown as open symbols in Fig.~\ref{fig:ft_a_ttot}a), while a fit (not shown) of these indentation data to a poroelastic contact model developed elsewhere~\cite{delavoipiere2016} provides a poroelastic time $\tau=19$~s. This shows that, in the low $Pe$ regime, the transient increase in $a(t)$ is uniquely dictated by the normal loading, independently of the imposed sliding. Hence, the contact relaxes to its steady state value $a_0$, which corresponds to the equilibrium indentation at $Pe=0$, with a kinetics that is independent on both static contact time and sliding velocity, or P\'eclet number as long as $Pe \leq 1$. These conclusions also apply to friction force relaxation dynamics.\\
\indent We now turn to transient sliding at $Pe$ larger than unity. Fig.~\ref{fig:ft_a_pe_gt_1} shows the time dependence of $a$ and $F_t$ for $F=200$~mN and a sliding velocity $v=45\: \mu$m~s$^{-1}$ which, considering the resulting contact sizes, corresponds to a $Pe$ of about 6. Here again, a varying static contact time is imposed before lateral motion is initiated. As shown in Fig.~\ref{fig:ft_a_pe_gt_1}, $F_t(t)$ and $a(t)$ in this P\'eclet regime are either increasing or decreasing functions towards the steady-state value depending on whether initial contact radius is smaller or greater than steady-state value $a_s$. The two situations are illustrated in Fig.~\ref{fig:schem_images}c (last two rows) where images of the contact are shown. For $t_{sta} =$ 1 s, the initial contact is smaller than in steady state. For $t_{sta} =$ 40 s, the opposite situation is observed. As compared to situations with $Pe \lesssim 1$, the steady state value of the contact radius $a_s$ is decreased from 138~$\mu$m to 128~$\mu$m. In this high $Pe$ regime, we have previously shown~\cite{delavoipiere2018} that the poroelastic flow in the gel across the typical length $a_0$ is slow compared to the sliding motion, which results in a pore pressure imbalance between the leading and trailing edges of the contact. This pressure imbalance due to advection generates a net lift force on the sphere. As a consequence, the contact size $a_s$ in steady state is smaller than the equilibrium contact size $a_0$ as shown in Fig.~\ref{fig:schem_images}c (second column). In addition, the contact is no longer symmetric: whereas the leading edge remains circular with contact radius denoted $a$ smaller than $a_0$, the trailing edge recedes inward. Such a contact asymmetry was effectively observed~\cite{delavoipiere2018} during steady state at $Pe>10$. Here, it was not clearly evidenced within the experimental resolution of the optical observations for the considered $Pe$ range (experiments at $Pe$ greater than about 10 systematically resulted in cavitation processes which fall out of the scope of the present study). Altogether, at P\'eclet number larger than unity, the transient relaxation of both $a$ and $F_t$ may result in either an undershoot, or an overshoot, that is, a friction peak.\\ 
\begin{figure}[!ht]
	\centering
	\includegraphics[width=0.9 \columnwidth]{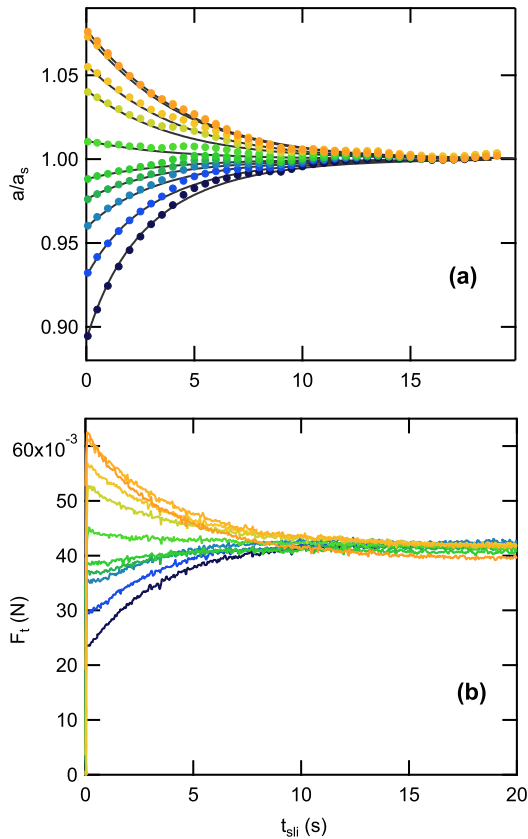}
	\caption{Contact radius $a$ normalized by steaty-state contact radius $a_s$ (a) and friction force $F_t$  (b) as a function of sliding time $t_{sli}$ during lateral motion at imposed velocity $v=45 \: \mu$m~s$^{-1}$ and applied normal force $F=200$~mN. Sliding motion is initiated after various static contact time $t_{sta}$. From bottom to top: 0.8, 1.8, 3, 4, 4.7, 6.9, 12.1, 16.9, 29.9 and 51.9 s. $Pe \approx 6$. Solid lines in (a) correspond to fits according to  Eqn~(\ref{eq:lift:dyn1}) with a characteristic time $\tau_s=7$ s. }
	\label{fig:ft_a_pe_gt_1}
\end{figure}
\indent We now examine the relationship between friction force and contact area. In Fig.~\ref{fig:fig6_ftfn_aa0_fn}, the friction force $F_t$ has been normalized by its steady-state value $F_s$ and is reported as a function of the normalized contact radius $a/a_s$, where $a_s$ is the steady state value of the contact radius. The normalization by $F_s$ is chosen to disregard the dispersion of friction force measurements. Experimental data correspond to normal forces ranging from $100$ to $500$~mN, sliding velocities ranging from 1 to 45~$\mu$m~s$^{-1}$ and static contact times ranging from 2 to 60~s. In Fig.~\ref{fig:fig6_ftfn_aa0_fn}, the regime $Pe<1$ corresponds to $a/a_s<1$, in that case $a_s \equiv a_0$, while for $Pe>1$ $a/a_s$ can be larger than unity. Although the experimentally achievable dynamics in $F_t/F_s$ and $a/a_s$ is restricted, the log-log plots in Fig.~\ref{fig:fig6_ftfn_aa0_fn} tend to indicate that $F_t(a/a_s)$ follows a power law with a velocity- and load-independent exponent close to $9/2$.\\
\begin{figure}[!ht]
	\centering
	\includegraphics[width=1 \columnwidth]{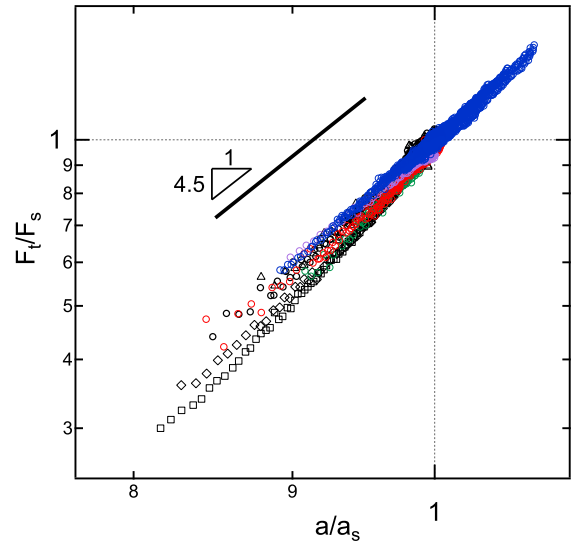}
	\caption{Normalized friction force $F_t/F_s$ versus normalized contact radius $a/a_s$,  where $F_s$ and $a_s$ are the steady-state friction force and contact radius, respectively. Blue: $v=45 \: \mu$m~s$^{-1}$; purple: $v=10 \:\mu$m~s$^{-1}$; black: $v=5 \:\mu$m~s$^{-1}$  ; red: $v=3 \:\mu$m~s$^{-1}$; green: $v=1 \:\mu$m~s$^{-1}$. $\square$: $F=500$~mN, $\diamond$: $F=300$~mN, $\circ$:  $F=200$~mN, $\triangle$:  $F=100$~mN. Associated $Pe$ numbers range from 0.15 to 6. Data correspond to static contact times ranging from 1~s to 60~s.}
	\label{fig:fig6_ftfn_aa0_fn}
\end{figure}
%
\indent In both the high and low P\'eclet regimes, a striking feature of the transient friction is therefore that the time-dependence of the friction force $F_t$ is entirely dictated by the time-dependence of the contact radius $a$, independently on the value of the normal force and of the sliding velocity. Moreover, when $Pe<1$, the kinetics of $a(t)$ is unaffected by the lateral displacement, thus suggesting that poroelastic flows induced by the normal loading and lateral displacement are fully decoupled. In what follows, we rationalize these observations from the extension of a previously developed poroelastic contact model~\cite{delavoipiere2018} to transient sliding situations. Within the limits of a thin film approximation, we first derive the expression of the pore pressure distribution during transient sliding. From this equation, the time-dependence of the contact radius can be accounted for. Under the assumption that frictional force arises solely from viscous dissipation within the pore network, an estimate of $F_t$ is also provided.\\
%
\section*{Contact model}
%
We consider a rigid sphere with radius $R$ indenting a thin layer of hydrogel with initial thickness $e_0$ resting on a substrate as depicted in Fig.~\ref{fig:schem_images}a. Both the sphere and the substrate are considered as rigid bodies. A normal force $F$ is imposed to the sphere, as well as a steady velocity $v$ in the direction $x$ parallel to the hydrogel layer. The moving coordinate system attached to the center of the contact is denoted $(x,y,z)$ ($x$ = 0 at the apex of the sphere) and $X$ is the position of a material point in the gel with respect to a fixed coordinate system. Along the lines of its steady-state formulation~\cite{delavoipiere2018}, our poroelastic model can account for the development of non circular contact shapes at high $Pe$. Accordingly, the contact between the sphere and the gel layer is described by a contact line defined by:
\begin{equation}
r=\rho(\theta), \text{with}\:\: \rho(\theta)=\rho(-\theta)
\end{equation}
in cylindrical coordinates $(r,\theta,z)$ attached to the sphere, corresponding to the $(x,y,z)$ coordinate system. We will next determine the stress field and the friction force in the transient regime where the sphere simultaneously indents the gel layer and slides past it.\\
The film thickness being much smaller than the contact size, we neglect the shear deformation within the layer, and the normal strain $\epsilon=(e_0-e)/e_0$ in the film reduces to:
\begin{eqnarray}
\epsilon (r,t)=\left\{\begin{matrix} 
\left( a^2-r^2 \right)/(2Re_0) & \left |r \right |<\rho\\ 
0  & \left |r \right |>\rho
\end{matrix}\right. 
\label{eq:epsilon}
\end{eqnarray}
where $a$ is the contact radius at the intercept with the $x$-axis as shown in Fig.\ref{fig:schem_images}b. In the frame of the gel with coordinate $(X,Y)$, the water flux $J$ within the hydrogel layer obeys the volume conservation equation: $ \mathop{\rm div} J-\left.{\frac{\partial \epsilon }{\partial t}}\right| _{X,Y}=0$. The second term also writes $\left.{\frac{\partial \epsilon}{\partial t}}\right | _{X,Y} = \left.{\frac{\partial \epsilon}{\partial t}}\right | _{x,y}-v\frac{\partial \epsilon }{\partial x}$ so that, in the frame of the sphere, the conservation equation writes:
\begin{align}
\mathop{\rm div} J&=\left.{\frac{\partial \epsilon}{\partial t}}\right | _{x,y} -v\frac{\partial \epsilon }{\partial x}\\
&=\left.{\frac{\partial \epsilon}{\partial t}}\right | _{r,\theta}-v\cos \theta \frac{\partial \epsilon }{\partial r}
\label{eq:conservation}
\end{align}
the strain $\epsilon$ within the contact being independent of $\theta$. The first term on the right hand side describes the indentation of the sphere in the gel, and the second term is the advective term. Following the poroelastic theory by Biot~\cite{biot1955,hui2005}, the normal stress $\sigma$ is assumed to be the sum of two terms: the elastic stress in the polymer network and the pressure $p$ of water in the pores:
\begin{equation}
\sigma(r,\theta,t)=\tilde{E} \epsilon (r,\theta,t) + p(r,\theta,t)
\label{eq:sigma}
\end{equation}
where $\tilde{E}$ is the uniaxial compression modulus of the drained polymer network
\begin{equation}
\tilde{E}=\frac{2G\left(1-\nu\right)}{1-2 \nu}
\label{eq:oedometric}
\end{equation}
where $G$ is the shear modulus and $\nu$ is the Poisson's ratio of the drained network. The pore pressure field is governed by Darcy's law:
\begin{eqnarray}
\overrightarrow{J}=-\kappa \overrightarrow{\nabla}p
\label{eq:darcy}
\end{eqnarray}
where $\kappa=D_p/\eta$ with $D_p$ the permeability of the gel network and $\eta$ the viscosity of the solvent. 	
Combining eqns~(\ref{eq:epsilon}), (\ref{eq:conservation}), and (\ref{eq:darcy}), the equation ruling the pore pressure field in transient regime writes:  
\begin{eqnarray}
\Delta p = -\frac{\dot{a}a}{e_0R\kappa}-\frac{vr \cos\theta}{e_0 R\kappa}
\label{eq:diff:pore_pressure}
\end{eqnarray}
with $\dot{a}$ the time derivative of the contact size. With no advection, namely $v=0$, we recently showed~\cite{delavoipiere2018} that the contact radius $a_0$ at equilibrium ($\dot{a}=0$) is determined by the normal force $F$, the hydrogel drained modulus $\tilde{E}$ and the geometrical parameters $e_0$ and $R$: 
\begin{equation}
a_0= \left(\frac{4Re_0F}{\pi\tilde{E}}\right)^{1/4}
\label{eq:a0}
\end{equation}
In addition, starting from an out of equilibrium contact size $a \neq a_0$, the contact relaxes to indentation equilibrium within the poroelastic time $\tau$ given by:
\begin{equation}
\tau=\frac{a_0^2}{4\kappa\tilde{E}}
\end{equation}
From this, all variables can be recast in non dimensional form:
\begin{align}
\overline{r} &\equiv \frac{r}{a}\\
\overline{a} & \equiv \frac{a}{a_0}\\
\overline{\dot{a}} &\equiv \frac{\dot{a}}{a_0}\tau\\
\overline{p} &\equiv \frac{p}{2F/\pi a_0^2}\\
\overline{\sigma}  &\equiv \frac{\sigma}{2F/\pi a_0^2}\\
Pe^0 &= \frac{\tau v}{a_0}
\end{align} 
so that the pore pressure eqn~(\ref{eq:diff:pore_pressure}) and the contact stress equation~(\ref{eq:sigma}) become:
\begin{equation}
\frac{1}{\overline{r}}\frac{\partial}{\partial \overline{r}}\left( \overline{r} \frac{\partial \overline{p}}{\partial \overline{r}}\right)+\frac{1}{\overline{r}^2}\frac{\partial ^2 \overline{p}}{\partial \theta^2}  
+8\overline{a}^3Pe^o \overline{r} \cos \theta=-8\overline{\dot{a}}\overline{a}^3
\label{eq:diff:pore_pressure_adim}
\end{equation}
\begin{equation} 
\overline{\sigma} =\overline{p}+(1-\overline{r}^2)\overline{a}^2
\label{eq:sigmaadim}
\end{equation}
Eqn~(\ref{eq:diff:pore_pressure_adim}) is a second order differential equation in $(r,\theta)$ with a time-dependent constant on the right hand side. The problem is closed by two conditions : at the contact line, the flux vanishes over a typical distance set by the film thickness, which is small compared to the contact radius. From this, the pore pressure is taken to be zero at the contact edge. Second, the total normal force is equal to the integral of the normal stress over the contact area: $F=\int \int \sigma rdrd\theta$. These boundary conditions write:
\begin{equation}
\overline{p}(r =\rho, t)=0
\label{eq:CL_pressure}
\end{equation}
\begin{equation}
\int\int_A \overline{\sigma}\overline{r}d\overline{r}d\theta = \frac{\pi}{2\overline{a}^2}
\label{eq:CL_cont_stress}
\end{equation}
A solution of the homogeneous equation derived from Eqn~(\ref{eq:diff:pore_pressure_adim}) verifying the first boundary condition writes:
\begin{equation}
\overline{p}^o(\overline{r},\theta,t)=\overline{a}^3Pe^0\left(\overline{r}\cos\theta(1-\overline{r}^2)+ \sum _{n=0}^{\infty}\overline{\alpha}_n \overline{r}^n \cos n\theta \right)
\label{eq:po:adim}
\end{equation}
so that the general solution of the differential Eqn~(\ref{eq:diff:pore_pressure_adim}) is:
\begin{equation}
\begin{split}
\overline{p}(\overline{r},\theta,t) = \overline{a}^3\left(Pe^0\overline{r}\cos\theta+2\overline{\dot{a}}\right)(1-\overline{r}^2)  \\ 
+ \overline{a}^3Pe^0\sum _{n=0}^{\infty}\overline{\alpha}_n \overline{r}^n cos n\theta
\end{split}
\label{eq:p:adim}
\end{equation} 
The pressure field is simply the superimposition of the pore pressures due to (i) the flow advected by the sliding of the gel with respect to the indenting sphere, and (ii) the flow forced by the indentation of the sphere into the gel layer (or its lifting out of it).\\
\indent When $v\neq0$ and $\overline{\dot{a}}=0$, eqn~(\ref{eq:p:adim}) reduces to the expression for pore pressure distribution under steady-state sliding which was discussed in a previous publication~\cite{delavoipiere2018}. When $Pe<1$, we showed that advection is slow enough so that the degree of indentation of the sphere into the hydrogel film is not modified by advection. The steady-state contact is circular, with a contact radius equal to $a_0$ and the sum term in  eqn~(\ref{eq:p:adim}) vanishes. When $Pe>1$, the contact radius $a_s$ in steady state is smaller than the equilibrium contact radius $a_0$ and the contact is no longer symmetric as a result of the lift force generated by the pressure imbalance across the contact. Note that, as depicted in Fig.~1b, $a_s$ is the contact radius along the positive x-axis. In this regime, contact asymmetry is accounted for by the non vanishing $\overline{\alpha}_n$ coefficients in eqn~(\ref{eq:p:adim}).\\
%
\indent In what follows, we derive the expression for the friction force under the assumption that frictional energy dissipation arises solely from the viscous dissipation associated with the pressure-driven water flow within the porous gel network. We thus neglect any dissipative mechanisms at the glass/gel interface arising, as an example, from adhesive forces. This assumption is supported by the previous observation that poroelastic indentation of the PDMA hydrogel film is perfectly described by our poroelastic model with no need to incorporate adhesive effects~\cite{delavoipiere2016}. Moreover, we assume that during sliding, energy is dissipated by the pore pressure term associated to advection, $\overline{p}^o$ (Eq.~\ref{eq:po:adim}), while the pore pressure term associated to indentation is balanced by the work done by the normal force. As for steady state sliding, the calculation is thus based on the determination of the energy dissipated by the poroelastic flow forced by advection, denoted $W$, per unit advance $\Delta$ of the sphere along the sliding direction $x$. Dissipation arises from successive squeezing and re-swelling of the gel network at the leading and trailing edges of the contact, respectively, so that the absolute value of the product $\left |p^o(x,y) x \right |$ is taken.
\begin{equation}
\frac{W}{\Delta}=\frac{1}{R}\int\int_A \left |p^o(x,y) x \right |dxdy
\label{eq:work0}\\
\end{equation}
Eqn~(\ref{eq:work0}) can be rewritten as
\begin{equation}
\frac{W}{\Delta} =I \frac{2Fa_0}{\pi R}Pe^0\overline{a}^6
\end{equation}
where
\begin{equation}
\begin{split}
I=\int\int_{\overline{A}} \Bigg| \overline{r}^2\cos\theta \Big( \overline{r}\cos\theta(1-\overline{r}^2) \\
+\sum _{n=0}^{\infty}\overline{\alpha}_n \overline{r}^n \cos n\theta \Big) \Bigg| d \overline{r} d\theta
\label{eq:I}
\end{split}
\end{equation}
and $\overline{A}$ is the contact area normalized by $A_0=\pi a_0^2$. The prefactor $I$ is a non dimensional factor that depends on the exact shape of the contact but not on its size. For circular contacts, all $\overline{\alpha}_n$ are zero, $\overline{r}$ varies between 0 and 1, and the prefactor $I$ reduces to $I=\pi/12$.\\
Following a fracture mechanics argument,\cite{delavoipiere2018} we assume that most of the poroelastic dissipation occurs close to the trailing and leading edges of the contact which can be viewed as advancing and receding cracks, respectively. The energy $\mathcal{G}_{poro}$ needed to drive these cracks is supposed to arise solely from poroelastic dissipation over the crack length
\begin{equation}
\mathcal{G}_{poro} \equiv \frac{1}{\pi a }\frac{W}{\Delta}
\end{equation}
Using a scaling argument, the frictional stress $\sigma_f$ is related to $\mathcal{G}_{poro}$ by
\begin{equation}
\sigma_f \approx \sqrt{\frac{E*}{e_0}\mathcal{G}_{poro}}
\end{equation}
Taking $E^* \approx \tilde{E}/3$ and recalling that $F=\pi\tilde{E}a_0^4/4Re_0$, we compute the friction force as $F_t=A\sigma_f$ where $A$ is the area of the possibly non circular contact.
\begin{equation}
\frac{F_t}{F} \approx \frac{4}{\sqrt{6 \pi}}\frac{A}{A_0}\overline{a}^2\sqrt{I}\sqrt{Pe^0}\sqrt{\overline{a}}
\label{eq:friction:general}
\end{equation}
where $A_0=\pi a_0^2$ is the area of the circular contact area corresponding to $Pe<1$ or indentation equilibrium with no sliding. Both $A$ and $I$ can be calculated from a knowledge of the actual contact shape.\\

\section*{Discussion}
%
We now discuss the time changes in contact shape in the transient regime for $Pe<1$ and $Pe>1$. Then, we describe how the friction force during transient sliding is dictated by the contact size for both P\'eclet regimes.
\subsubsection*{Time evolution of contact radius}
\- For $Pe<1$, contacts remain circular both in the transient regime when the sphere indents the gel layer and in steady state. For circular contacts, the boundary condition Eq.~\ref{eq:CL_pressure} and Eq.~\ref{eq:p:adim} imply that $\overline{\alpha}_n=0$ for all $n$ values. Also, the condition on the total normal contact stress (eqn~(\ref{eq:CL_cont_stress})) gives the time-dependence of the contact radius in the transient regime when the sliding starts before indentation has come to a halt. The latter writes:
\begin{equation}
\overline{a}^4\left(1+2\overline{\dot{a}}\overline{a}\right)=1
\label{eq:indentation:eq:diff}
\end{equation}
Solving the differential equation for $a$ yields, with initial condition $a(t=0)=0$:
\begin{eqnarray}
\overline{t}=-\overline{a}^2 + Argth \left(\overline{a}^2\right)
\label{eq:indentation:dyn}
\end{eqnarray}
where $\overline{t}=t/\tau$ and $Argth(x)=1/2\ln((x+1)/(x-1))$. This expression corresponds to the indentation equation that was used to fit the experimental data with a zero velocity $v$ in Fig.~\ref{fig:ft_a_ttot}a, in order to compute the poroelastic time $\tau$.  The  sliding movement at velocity $v$ does not influence the indentation dynamics in the low P\'eclet regime. In Fig.~\ref{fig:ft_a_ttot}a, contact radius data with different static contact time collapsed on a single master curve corresponding to the indentation curve $a(t)$ measured in static conditions. This observation justifies the expression of the pressure field as the superimposition of the pore pressures respectively due to advection and to indentation of the sphere into the gel. Both theory and experiments thus suggest that the time-dependence of the contact radius in low P\'eclet regime is only controlled by the normal loading regardless of the imposed lateral displacement. \\
For $Pe>1$, the sum term in the expression of pore pressure (eqn~(\ref{eq:p:adim})) is no longer vanishing as the contact shape is predicted to be non circular. As detailed in Supporting Information (SI), eqns~(\ref{eq:sigmaadim},\ref{eq:p:adim}) still can be solved numerically with the appropriate boundary conditions (Eqns.~(\ref{eq:CL_pressure}-\ref{eq:CL_cont_stress}) in order to determine the unknown $\alpha_n$ coefficients and the associated contact shape during the transient regime. Consistently with the experimental observations, these numerical simulations indicate that contact asymmetry remains very limited for the $Pe$ range under consideration.\\
We can take therefore advantage of the observation that, within the investigated $Pe$ range, the contact shape is evolving with very limited asymmetry to derive an approximate solution for $a(t)$ and $F(t)$. For $Pe>1$, the steady state contact size $a_s$ is smaller than the static size $a_0$ so that, as a starting point, we consider that an effective load $F_{eq}$ smaller than the imposed normal load $F$ is applied to the contact as a result of the lift force $F_{lift}$ generated by the pore pressure imbalance. In this framework, the effect of the advected flow at velocity $v$ reduces to the lift force and is accounted for in the equivalent normal load. Here $F_{eq}$ will be assumed to be constant over time and immediately reached as soon as sliding starts.
\begin{equation}
F_{eq}=F-F_{lift}
\end{equation}
Under the assumption that the contact remains nearly circular, $F_{eq}$ can be estimated from the steady state contact radius $a_s$ using eqn~(\ref{eq:a0}) as
\begin{equation}
F_{eq} \approx\frac{\pi \tilde{E}}{4Re_0}a_s^4
\label{eq:ass_Feq}
\end{equation}
Further calculations detailed in the appendix show that when normalized by its steady-state value, the contact radius $\tilde{a}=a/a_s$ follows a differential equation similar to the indentation equation eqn~(\ref{eq:indentation:eq:diff}) provided that the times are normalized by $\tau(a_s)=\tau a_s^2/a_0^2$, yielding $\tilde{t}=t/\tau(a_s)$. In the transient regimes, the sphere may either further indent the gel film, or be lifted out of it depending on the relative values of the contact radius at initial sliding time ($a_{ini}$) and at steady state ($a_{s}$). The initial value of the contact radius $a_{ini}$  is either larger or smaller than the steady state value $a_s$ depending on the static contact time.\\
The associated poroelastic time $\tau(a_s)=\tau a_s^2/a_0^2$ characterizes the poroelastic drainage towards the new equilibrium contact size $a_s$. Consistently with the numerical simulations by Qi~\textit{et al}~\cite{qi2020}, this characteristic time associated to the transient regime is thus shorter than that of normal indentation when $Pe>1$, as $a_s < a_0$. The equation describing the time-dependence of the contact radius is:
\begin{eqnarray}
\tilde{t}-\tilde{t}_{sta}=\tilde{a}_{ini}^2- \tilde{a}^2 + \frac{1}{2} \ln \left( \frac{\tilde{a}^2+1}{\tilde{a}_{ini}^2+1} \frac{\tilde{a}_{ini}^2-1}{\tilde{a}^2-1}\right)
\label{eq:lift:dyn1}
\end{eqnarray}
Eqn~(\ref{eq:lift:dyn1}) was used to fit the experimental data of contact radius presented in Fig.~\ref{fig:ft_a_pe_gt_1} with $\tau(a_s)$ as fitting parameter. Remarkably, a single value of characteristic time $\tau_{fit}$ could be used to fit all the data to eqn~(\ref{eq:lift:dyn1}) whatever the initial value of contact radius (see Fig.~\ref{fig:ft_a_pe_gt_1}a). The associated characteristic time $\tau_{fit}=7$~s differs from our prediction giving $\tau(a_s)=\tau a_s^2 / a_0^2 = 17$~s. This discrepancy is attributed to the simplifying assumption of constant $F_{eq}$ and circular contact of the model (see SI). Nevertheless, our approximate model captures three attributes of transient sliding at high P\'eclet numbers in good agreement with experimental data. First, relaxation towards a steady-state is faster than in the low P\'eclet regime : $\tau(a_s)<\tau$. Then, two behaviors of undershoot ($a_{ini}<a_s$) or overshoot ($a_{ini}>a_s$) are predicted and a unique characteristic time describes both regimes. In addition, except from the value of $\tau(a_s)$, the time-dependence of the contact radius is remarkably described by the model.\\
\subsubsection*{Friction force}
\- Friction force was considered to result solely from viscous dissipation within the porous network, neglecting interfacial contributions. For $Pe<1$, the transient contact shape is circular at all times and the friction force can be easily derived from eqn~(\ref{eq:friction:general}) where the term $I$ reduces to $\pi/12$:
\begin{align}
\frac{F_t}{F}\approx \frac{\sqrt{2}}{3}\sqrt{Pe^0}\overline{a}^{9/2}
\label{eq:indentation:force}
\end{align}
This demonstrates that for circular contacts, the friction force is entirely determined by the transient contact radius $a=\overline{a}a_0$.\\
The scaling of the friction force with the contact radius was tested for various P\'eclet values in Fig.~\ref{fig:fig6_ftfn_aa0_fn} where the friction force $F_t$ is normalized by its steady state value $F_s$ and reported as a function of $\tilde{a}=a/a_s$. In this framework,  dividing eqn~(\ref{eq:friction:general}) by its steady state expression yields :
\begin{align}
\frac{F_t}{F_s} \approx \frac{A}{A_s}\frac{a^2}{a_s^2}\frac{\sqrt{I}}{\sqrt{I_s}}\frac{\sqrt{a}}{\sqrt{a_s}}
\label{eq:friction:normalized}
\end{align}
All the experimental data collapsed on a mastercurve with a slope close to $9/2$ for P\'eclet values below and over 1. This observation was expected for $ Pe<1$ where $a_s=a_0$ and  eqn~(\ref{eq:friction:normalized}) yields:
\begin{align}
\frac{F_t}{F_s}\approx \left(\frac{a}{a_s}\right)^{9/2}
\end{align}
More surprisingly, the scaling of friction force in $\tilde{a}^{9/2}$ still holds when $Pe>1$. The contact asymmetry at large P\'eclet, represented by the terms $I$ and $A$, appear to have little influence on the friction force expression as a function of the contact radius whereas the same slight asymmetry of the contact had a strong impact on the time-dependence of the contact radius. These experimental observations along with theoretical predictions highlight a very singular feature of the transient sliding of a sphere on a hydrogel layer : in either a low or a large P\'eclet regime, the evolution of the normalized friction force $F_t(t)/F_s$ with time is entirely set by the evolution of the contact radius $a(t)/a_s$ for a given sliding velocity.
%
\section*{Conclusion}
%
Contact experiments on thin hydrogel films in water evidenced the dominant contribution of poroelastic flow to the transient frictional response involved in the onset of sliding. Starting from the contact at rest, poroelastic dissipation within the hydrogel network was observed to result in a transient regime which extends well beyond the achievement of a full sliding condition at the interface. During this transient regime, the dynamics of the frictional force is strongly correlated to the progressive adjustment of the contact radius to its steady-state value as a result of pressure-induced water flow within the hydrogel pores with very limited contribution from interface dissipation. Depending on the relative values of the initial and steady-state contact sizes and thus on the static contact time before sliding, the transient friction force shows either undershoots or overshoots: the latter case only occurs at P\'eclet numbers larger than 1, where friction peaks can be observed as a result of poroelastic flow.\\ 
The development of a poroelastic contact model using a thin film approximation allowed to capture the main features of the transient regime, especially the occurrence of the so-called static friction peaks. Furthermore, we show that the ratio of the transient to steady state friction force, $F_t(t)/F_s$, uniquely depends on the time-dependence of the normalized contact radius $a(t)/a_s$, through a power law with exponent 9/2, whatever the applied normal force or imposed velocity. This is a very unique and remarkably simple result. Although qualitatively similar transient regimes could be anticipated, as an example, within lubricated contact with viscoelastic substrates as a result of the time-dependent mechanical properties, their description would probably be much complicated by the wide distribution of the relaxation times. Here, a single characteristic poroelastic time can be ascribed to the transient.\\
Our poroelastic description of frictional transient regime could be straightforwardly extended to many situations relevant to practical applications where, as an example, fluctuations in the sliding velocity or contact load are involved. We also believe that our work sets the basis for studies of transient friction of hydrogels with a higher degree of complexity, where enhanced molecular interactions at the gel/slider interface would result in a coupling between poroelastic flow and interface dissipation.\\
\section*{Acknowledgments}
One of us (J. Delavoipi\`ere) is indebted to Ekkachai Martwong for his kind support during the synthesis of the hydrogels systems. The authors also wish to thank C.-Y. Hui for stimulating discussions. 
%
\appendix
\section*{Appendix}
\numberwithin{equation}{subsection}
\renewcommand{\thesubsection}{\Alph{subsection}}
\subsection{Derivation of the approximate $a(t)$ relationship for $Pe>1$}
Depending on the magnitude of the lift force, the pore pressure field $p_{eq}$ associated to $F_{eq}$ can be either a suction term (negative pressure at the contact center) or an overpressure (positive pressure) associated with an decrease or an in increase in contact radius, respectively. It writes
\begin{eqnarray}
p_{eq}(r,\theta,t)=\frac{2a\dot{a}}{8e_0 R\kappa}(a^2(t)-r^2)  
\end{eqnarray}
From  eqn~(\ref{eq:epsilon}) and eqn~(\ref{eq:sigma}), the total equivalent stress writes:
\begin{eqnarray}
\sigma_{eq}(r,t)=\tilde{E} \epsilon (r,t) + p_{eq}(r,t)\\
\sigma_{eq}(r,t)=\left(\frac{2a\dot{a}}{8e_0 R\kappa}+\frac{\tilde{E}}{2Re_0}\right)(a^2(t)-r^2)  \\
\sigma_{eq}(r,t)=\frac{\tilde{E}}{2Re_0}\left(\frac{a\dot{a}}{2\kappa\tilde{E}}+1\right)(a^2(t)-r^2)
\label{eq:sigmaeq}
\end{eqnarray}
The equivalent normal load writes:
\begin{eqnarray}
F_{eq}=\int \int \sigma_{eq} rdrd\theta\\
F_{eq}=\frac{\tilde{E}}{2Re_0}2\pi\frac{a^4}{8}\left(\frac{a\dot{a}}{\kappa\tilde{E}}+1\right)
\end{eqnarray}
Using eqn~(\ref{eq:ass_Feq}), we find that the contact radius should satisfy the following relationship during transient:
\begin{equation}
1=\frac{a^4}{a_s^4}\left(2\frac{a}{a_s}\frac{\dot{a}}{a_s}\tau(a_s)+1\right)
\end{equation}
where the poroelastic time $\tau(a_s)=\frac{a_s^2}{4\kappa\tilde{E}}$ characterizes the poroelastic drainage towards the new equilibrium contact size $a_s$.\\

This differential equation is supplemented with the following initial conditions:
\begin{align}
a(t=t_{sta})&=a_{ini}\\
a(t\rightarrow\infty)&=a_s
\end{align}
The problem can be recast in non dimensional form using the reduced contact size $\tilde{a}=\frac{a}{a_s}$ and reduced time $\tilde{t}=\frac{t}{\tau_s}$. It is formally identical to the indentation case, except initial contact size may be larger than steady state contact size. 
\begin{align}
\tilde{a}^4\left(2\tilde{\dot{a}}\tilde{a}+1\right)&=1\\
\tilde{a}(t=t_{sta})&=\tilde{a}_{ini}\\
\tilde{a}(t\rightarrow\infty)&=1
\end{align}
Solving the differential equation for $\tilde{a}$ yields:
\begin{eqnarray}
\tilde{t}-\tilde{t}_{sta}=\tilde{a}_{ini}^2- \tilde{a}^2 + \frac{1}{2} \ln \left( \frac{\tilde{a}^2+1}{\tilde{a}_{ini}^2+1} \frac{\tilde{a}_{ini}^2-1}{\tilde{a}^2-1}\right)
\label{eq:lift:dyn}
\end{eqnarray}
%
%
\bibliographystyle{rsc}

{10}

\newpage

\begin{center}
	\section*{Numerical resolution of the poroelastic contact model for $Pe>1$.}
\end{center}
We describe here the numerical resolution of Eqns~(19,23) and (35) for the contact pressure $\overline{\sigma}(r,\theta)$, the pore pressure field $\overline{p}(r,\theta)$,  and the friction force $F_t$, respectively, when $Pe>1$. In this Péclet regime, the contact is no longer circular as a result of the pore pressure imbalance generated by advection. An additional unknown of the poroelastic contact problem is thus the contact shape. Here, we make the assumption that the contact line has the general form $\overline{r}=\overline{\rho}(\theta)$ where $\overline{\rho}=\rho/a$ is an even function of $\theta$. Consistently with the experimental observation reported in Delavoipière~\textit{et al}~\cite{delavoipiere2018}, the contour line $\rho$ is assumed to be described by an ellipse for $|\theta| \geq\pi/2$ while the leading edge of the contact remains circular ($|\theta| \leq\pi/2$).
\begin{equation}
\begin{split}
r=a \:\:\: |\theta| \leq \pi/2 \\
r^2\cos^2 \theta+\frac{r^2\sin^2 \theta}{\zeta^2}=&a^2 \:\:\: |\theta| \geq\pi/2 
\end{split}
\end{equation}

which can be recast in a non dimensional form:

\begin{equation}
\begin{split}
\overline{r}=1 \:\:\: |\theta| \leq \pi/2 \\
\overline{r}^2\cos^2 \theta+\frac{\overline{r}^2\sin^2 \theta}{\zeta^2}=&1 \:\:\: |\theta| \geq\pi/2 
\end{split}
\label{eq:contact:line:shape}
\end{equation}
where $0<\zeta<1$ is a numerical parameter describing the contact asymmetry.\\
For each Péclet number $Pe^0$, the three unknowns $\overline{a}$, $\zeta$ and $\overline{\dot{a}}$ have to be determined as a function of the reduced time $\overline{t}=t/\tau$. The solution of the contact problem should comply to the following boundary conditions:\\

\begin{figure} [!ht]
	\centering
	\includegraphics[width=0.8\columnwidth]{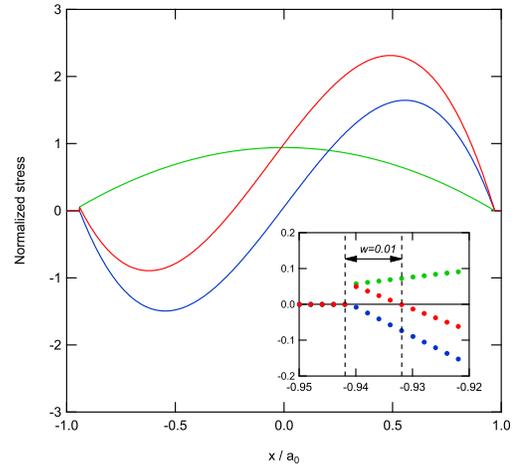}
	\caption{Profiles of the calculated pore pressure $\overline{p}$ (blue), contact stress $\overline{\sigma}$ (red) and elastic stress $\sigma_e=(1-\overline{r})\overline{a}^2$ (green) along the sliding direction $x$ ($y=0$). Inset: details of the profiles at the leading edge of the contact where $\overline{w}=0.01$ is the normalized width over which the contact stress $\overline{\sigma}$ remains positive from the edge of the contact.}
	\label{fig:si_fig1}
\end{figure}

\begin{enumerate}
	\item the pore pressure $\overline{p}(\overline{r},\theta)$ is vanishing at the contact edge ( cf eqn~(20)),
	\item the contact stress $\overline{\sigma}(\overline{r},\theta)$ should fulfill eqn~(21), i.e
	\begin{equation}
	\begin{split}
	&\int \int _A \overline{\sigma} dA=\frac{\pi}{2\overline{a}^2}
	\end{split}
	\end{equation} 
	where $A$ is the actual contact.
	\item We also enforce the condition that the contact stress $\overline{\sigma}$ cannot be negative at the edge of the contact, otherwise it will open: indeed, adhesion is not expected to be significant with the contact fully immersed within water, and adhesive forces are not accounted for in the model. However, negative (suction) contact stresses are allowed within the contact area provided that $\overline{\sigma}$ remains positive over a prescribed reduced length $\overline{w}=w/a_0$ at the edge of the contact (see Fig.~\ref{fig:si_fig1}). Here, we enforce the criterion $\overline{w} \geq 0.01$ which corresponds to the thickness of the hydrogel layer used in the experiments (i.e. $w/e_0 \approx 1$ ).
\end{enumerate}	
%
\subsection*{Steady-state sliding}
\begin{figure} [!ht]
	\centering
	\includegraphics[width=0.8\columnwidth]{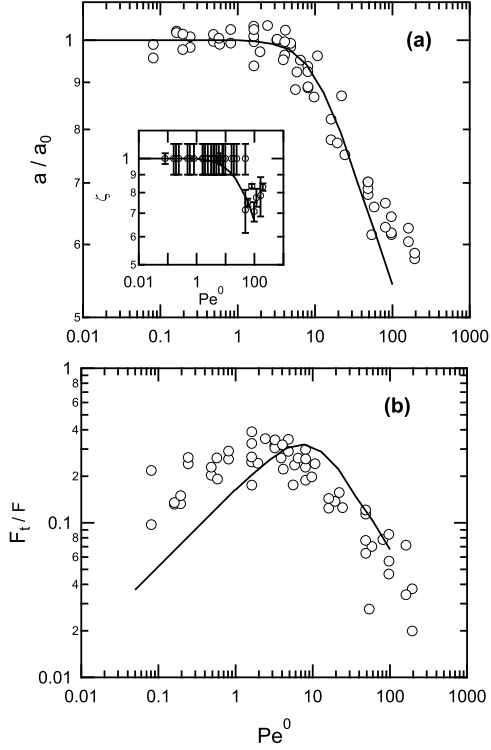}
	\caption{Numerical (solid line) and experimental (open symbols) (a) contact shape ($\overline{a}$,$\zeta$) and (b) reduced friction force $Ft/F$ as a function of $Pe^0$. Open symbols corresponds to experimental data for PDMA films taken from the present study and Delavoipière~\textit{et al}~\cite{delavoipiere2018}.}
	\label{fig:si_fig2}
\end{figure}

For steady-state sliding, $\overline{\dot{a}}=0$ and the problem reduces to the determination of $\overline{a}$ and $\zeta$. For each of the considered $Pe$ number, we start from the calculation of the pore pressure and contact stress fields for various combinations of $\overline{a}$ and $\zeta$ within the range $0.4<\overline{a}<1$ and $0.5<\zeta<1$. For each ($\overline{a}$, $\zeta$) doublet, we determine the set of $\alpha_n$ parameters ensuring $\overline{p}=0$ on the considered contact line defined by Eq.\ref{eq:contact:line:shape}. Once $\alpha_n$ are determined, $\overline{p}$ and $\overline{\sigma}$ can be calculated everywhere in the contact area using eqns~(19,23). Then, we search in the discretized ($\overline{a}$, $\zeta$) space the solution which obeys the second and third boundary conditions.\\
The calculated $\overline{a}(Pe^0)$ and $\zeta(Pe^0)$ solutions are shown in Fig.~\ref{fig:si_fig2}(a) together with some experimental data taken from Delavoipière~\textit{et al}~\cite{delavoipiere2018} and from the present study. In Fig.~\ref{fig:si_fig2}(b), the calculated $F_t/F$ curve has been shifted vertically to fit the experimental data as the expression of the friction force is derived from a scaling argument (eqn~(28)) with an unknown prefactor. The maximum in friction force occurs for a $Pe^0$ greater than unity ($Pe^0 \approx 8$) as a result of two competing effects (\textit{i}) the increase in viscous dissipation per unit film volume within the contact as $Pe^0$ is increased, (\textit{ii}) the decrease in the size of the contact, i.e. in the film volume affected by poroelastic flow.\\
For the highest $Pe$ value achieved in the transient experiments of this paper (i.e. $Pe=6$), the calculated contact asymmetry ($\zeta=0.966$) is very close to unity, in agreement with the experimental observation that the contact remains nearly circular within optical resolution. 
\subsection*{Transient sliding}
\begin{figure} [!ht]
	\centering
	\includegraphics[width=0.7\columnwidth]{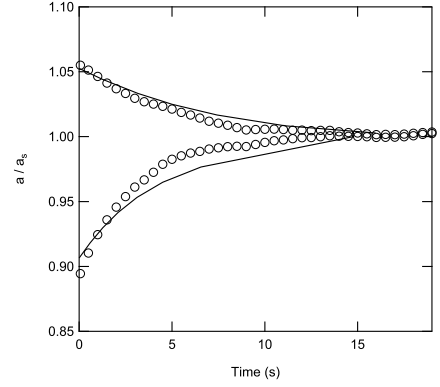}
	\caption{Numerical (solid lines) and experimental (open symbols) $a(t)$ kinetics for $Pe=6$ and two different values of the initial contact area ($a/a_s=0.89$,$a/a_s=1.05$).}
	\label{fig:si_fig3}%
\end{figure}

For transient sliding, along the same lines we calculate the contact stresses in a discretized ($\overline{\dot{a}}$,$\zeta$) space for the considered $Pe$ number and a discretized set of $\overline{a}$ values ranging from the initial ($\overline{a}_i$) to the steady-state ($\overline{a}_s$) contact radii. For each value of $\overline{a}$, and in the ($\overline{\dot{a}}$,$\zeta$) space, we determine the solution which satisfies the three above mentioned boundary conditions. The reduced time increment $\Delta \tau$ separating to successive contact radius values is determined as
\begin{equation}
\Delta \tau=\frac{\overline{a}_{n+1}-\overline{a}_{n}}{\overline{\dot{a}}_{n+1}}
\end{equation}
An example of the calculated $a(t)$ is shown in Fig.~\ref{fig:si_fig3} for $Pe=6$ and two different initial contact radii, together with the corresponding experimental data, and is found to be in fair agreement.
\subsubsection*{Equivalent force $F_{eq}$}
In order to provide an approximate solution for the transient at $Pe>1$, we made the assumption that an effective load $F_{eq}=F-F_{lift}$ smaller than the imposed load $F$ is applied to the contact as a result of the lift force $F_{lift}$ generated by the pore pressure imbalance. The effective load was approximated from the steady-state contact radius $a_s$ under the assumption that the contact remains nearly circular (eqn~(33)). It is therefore taken as constant during the transient.\\ 

Here, the numerical simulation allows to calculate the time-dependence of this effective load while taking into account the actual non-circular shape of the contact. Note that, consistently with the non-dimensional variables used in the paper, all forces are normalized by $2F/\pi$. The lift force arises from  the advective component of pore pressure (eqn~(22)). Hence, the normalized expression of the applied load $\overline{F}$, the lift force $\overline{F}_{lift}$, the effective load $\overline{F}_{eq}$ write:
\begin{eqnarray}
\begin{split}
\overline{F}&=\pi/2\\
\overline{F}_{lift}&=\int\int_A \overline{a}^2 \overline{p^o}\overline{r}d\overline{r}d\theta\\
\overline{F}_{eq}&=\int\int_A \overline{a}^2 (\overline{\sigma}-\overline{p^o})\overline{r}d\overline{r}d\theta
\end{split}
\end{eqnarray}
Using eqn~(23), $\overline{F}_{eq}$ simply writes:
\begin{equation}
\overline{F}_{eq}=\overline{a}^4\int\int_A \left(1-\overline{r}^2\right) \overline{r}d\overline{r}d\theta
\end{equation}
In the numerical model, the contact line is assumed to be circular at the leading edge and elliptic at the trailing edge (Eq.~\ref{eq:contact:line:shape}). Under this assumption, an analytical expression of $\overline{F}_{eq}$ can be derived as a function of $\overline{a}$ and $\zeta$:
\begin{equation}
\overline{F}_{eq}=\frac{\pi}{2}\overline{a}^4\frac{1}{2}\left(1+\frac{\zeta}{2}(3-\zeta^2)\right)
\label{eq:eff:load}
\end{equation}
In the limit of small contact assymetry, $\zeta$ is close to 1 and the effective load in Eq.\ref{eq:eff:load} amounts to the approximation made in eqn~(33): $\overline{F}_{eq}\sim\pi\overline{a}^4/2$.\\

\begin{figure} [!ht]
	\centering
	\includegraphics[width=0.7\columnwidth]{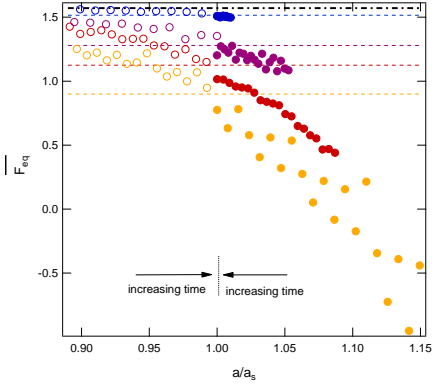}
	\caption{Calculated normalized equivalent force $\overline{F}_{eq}$ as a function of $a/a_s$ for various values of the Péclet number. Blue: $Pe=3$; Purple: $Pe=6$; Red: $Pe=10$; Yellow $Pe=15$. Open and filled symbols correspond respectively to initial values of $a/a_s$ smaller and larger than 1. The horizontal dotted lines correspond to $\overline{F}_{eq}=\pi \overline{a}_s^4/2$. The dash dotted line correspond to the normalized applied normal load $\overline{F}=\pi/2$.}
	\label{fig:si_fig4}%
\end{figure}

The equivalent force $\overline{F}_{eq}$ has been calculated for various $Pe$ numbers, either starting from the indentation equilibrium {\it i.e}. $a=a_0$ and $\overline{a}>1$ or starting from a value smaller than the steady state contact radius {\it i.e.} $\overline{a}<1$. In Fig.~\ref{fig:si_fig4}, the results were reported as a function of the time-dependent normalized contact radius $\tilde{a}=a(t)/a_s$ achieved at the various stages of the transient, where $a_s$ is the steady-state contact radius. It can be seen that, starting from $a_i$ smaller or larger than $a_s$, $\overline{F}_{eq}$ is respectively slowly decreasing or increasing toward the limit corresponding to $\overline{F}_{eq}\sim\pi\overline{a}_s^4/2$ (eqn~(33)) as $a$ reaches its steady state value ($\tilde{a}\rightarrow 1$) i.e. toward the frictional equilibrium. The amplitude of this change is enhanced when $Pe$ is increased. However, the hypothesis of constant effective load during the transient is acceptable for $Pe\leq6$: this supports the assumptions made in the derivation of an approximate analytical solution for the time variations of the contact radius $a$ (eqn~(34)).
\bibliographystyle{rsc}
	%

%
\end{document}